\documentstyle[twocolumn,prl,aps,epsf]{revtex}

\draft

\begin{document}

\twocolumn[\hsize\textwidth\columnwidth\hsize\csname
@twocolumnfalse\endcsname

\title{Surface and Image-Potential States on the $MgB_2(0001)$ Surfaces}

\author{V. M. Silkin$^1$, E. V.Chulkov$^{1,2}$, 
 and P. M. Echenique $^{1,2}$ }

\address
{$^1$Donostia International Physics Center (DIPC) and Centro Mixto
CSIC-UPV/EHU, 20018 San Sebasti{\'a}n, Spain\\
$^2$Departamento de F{\'\i}sica de Materiales, Facultad de
Ciencias Qu{\'\i}micas, Universidad del Pais Vasco/Euskal Herriko
Unibertsitatea, Apdo. $1072, 20018$ San Sebasti\'an/Donostia, Basque Country,
Spain}

\date{\today}

\maketitle

\begin{abstract}
We present a self-consistent pseudopotential calculation of surface and 
image-potential states on $MgB_2(0001)$ for both $B$-terminated ($B-t$) and 
$Mg$-terminated ($Mg-t$) surfaces. We find a variety of very clear surface 
and subsurface states as well as
resonance image-potential states n=1,2 on both surfaces. 
The surface layer DOS at $E_F$ is increased by 55\% at $B-t$ and by 90\% 
at the $Mg-t$ surface compared to DOS in the corresponding bulk layers.
\end{abstract}

]

The discovery of the superconductivity  in a simple metal polycrystalline compound $MgB_2$ with 
the critical temperature $T_c \sim$ 39 K \cite{naga2001} has generated an explosion
of research activity in studying the mechanism of the superconductivity
and properties of this compound \cite{bud'ko,finn,takahashi,rubio,karapetrov,giubileo,chen,sharoni,0102389,kortus,an,dolgov,hirsch,voelker,osborn,guinea,liu,gorshunov,jorg,saito,antropov,satta,taku,bohnen}. 
For instance, the superconductivity gap has been measured by both  bulk sensitive 
methods \cite{takahashi,gorshunov} and surface sensitive techniques. Compared to bulk
measurements
the surface sensitive experiments, namely scanning tunneling spectroscopy (STS)
\cite{rubio,karapetrov,giubileo,chen,sharoni} and point-contact experiments \cite{0102389},
give generally a smaller energy gap varying in the surface region \cite{giubileo,0102389}.
It may be caused by two effects: surface contaminations or/and desorder and by changing
the electronic structure at the surface. Qualitatively different STS spectra obtained 
by different groups on polycrystalline $MgB_2$ pellets and films reflect different 
surface contaminations and microstructure of the sample surfaces. However, for 
a single crystal a possible change of a high density of states (DOS) at the Fermi 
level, $E_F$, high phonon frequencies and strong electron-phonon interactions can 
also lead to a change of the energy gap and $T_c$ at the surface. 
Very recently two groups have announced obtaining single crystalline $MgB_2$ with edge
angles of 120 degrees \cite{jung,xu}. These works open up new prospects for experimental 
studies of the surface superconductivity in $MgB_2$.

Due to strong covalent interactions within 
B planes \cite{kortus,an,antropov,satta} the (0001)   
termination of $MgB_2$ is
supposed to be more  favorable. However nothing is known about atoms which form the topmost
layer of $MgB_2(0001)$. The study of the (0001) termination of other metal diborides
which also have crystal structures of the $AlB_2$ type have shown that some metal diborides
($TiB_2$, $HfB_2$) are terminated by metal atoms \cite{hayami,pesiss00} while the topmost
layer of $TaB_2$ is formed by the graphitic boron layer \cite{kasoprl98}. 
Here we report on {\it ab initio} calculations of the electronic 
structure  of the $MgB_2$(0001) surface for both types of termination.
In order to assess the effect of the surface relaxation on surface states
we have computed the surface electronic structure for the ideally bulk terminated crystal
as well as for contracted and expanded by 6\% the first interlayer spacing.

The bulk electronic structure of $MgB_2$ \cite{kortus,an,antropov,satta} 
leads to an unconventional bulk states
projection with very wide absolute and symmetry energy gaps (Fig. 1) which
support a variety of surface states and give an additional contribution to crystal 
reflectivity at energy interval just below the vacuum level where resonance image-potential 
states arise.
The surface and image states are of crucial importance for the description of the surface dynamical 
screening, electron (hole) excitations, and superconductivity at $MgB_2$ surfaces.
We show that for the $Mg$-terminated ($Mg-t$) surface the surface states contribution nearly 
doubles the surface DOS at $E_F$ compared to the bulk $Mg$ layer DOS. For the $B$-terminated 
($B-t$) surface the surface state contribution increases the surface DOS at $E_F$ by 55\%
compared to that in bulk. Special attention in the paper is focused on image potential 
states. We find that $Mg$- and $B$-layers possess distinct reflectivity that leads to 
different localization of image state wave function in bulk region. 

The calculations of charge density have been performed within 
the self-consistent 
local density-functional 
plane-wave method by using a supercell of 13 atomic layers and 7 layers 
of vacuum \cite{description}. This supercell is big enough to ensure a good description 
of both surface and bulk states. 
Experimental values of lattice constants $a$= 5.8317 a.u. and $c$= 6.6216 a.u. used 
in the evaluation have 
been taken from Ref.\cite{naga2001}.
The 13 layer slab representing the $B-t$ ($Mg-t$) surface consists of 7 $B$($Mg$) layers
alternating with 6 $Mg$($B$) layers.

As the LDA potential does not describe the correct asymptotic potential behavior
in the vacuum region we modify it by retaining the self-consistent
LDA one for $z<z_{im}$, where $z_{im}$ is the image plane position, and replacing it 
in the vacuum region for $z>z_{im}$  
by $V(z)=\{exp[-\lambda (z-z_{im})]-1\}/[4(z-z_{im})]$. The damping parameter $\lambda$
is a function of $(x,y)$ and is fixed by the requirement of continuity of 
the potential at $z=z_{im}$ for each couple of $(x,y)$.
With the use of the self-consistent charge density obtained for a 13-layer slab
we have constructed charge density for a 31-layer slab inserting 18 bulk layers
into the center of the slab. Vacuum space was increased from 7 layers to 21 ones.
This vacuum interval is enough to accurately describe the n=1 and 2 image states.
Finally the LDA potential was generated for this new supercell with a correct image
tail in vacuum space. 

In Figs.1a and 1b we show the calculated projection of bulk band 
structure onto the surface 
Brillouin zone together with surface states for $B-t$ and $Mg-t$
surfaces, respectively.
The light grey areas show the $\pi$ projected bulk states and the grey ones
indicate the $\sigma$ states. 
A remarkable feature of the bulk states projection is the presence of two wide
absolute energy gaps. The lower gap separates the $s$-bulk bands and $p_z$ 
bands of boron,
the upper gap crossed by the $p_{x,y}$ bulk bands of $B$ is located in the vicinity 
of the Fermi level, $E_F$.
The $B-t$ surface (Fig.1a) has 4 surface states strongly localized in 
the topmost boron layer and 3 subsurface states. All these states show energy dispersion
which repeats that of bulk bands. Two surface states degenerated 
at the $\overline{\Gamma}$ point
are of the $p_{x,y}$ symmetry ($\sigma$ states). They split off from bulk states
of the same symmetry by 0.45 eV and have energy of 1.23 eV relative to $E_F$. 
Their charge density is completely localized in the topmost layer (Fig.2). 
One can consider these states as two-dimensional quantum-well
states due to their extremely strong localization in the z direction: they decay
into the bulk much faster than do conventional surface states which are characterized by
a smooth exponential decay.  
Another surface state  with energy of -2.74 eV is of the $p_{z}$ symmetry
($\pi$ state), 75\% of the state is concentrated in the three surface atomic layers and 
in the vacuum region (Fig.3). The lower surface state  
is of $s$ symmetry and
splits off from bulk states by 0.4 eV, 70\% of the state is localized in the topmost layer. 
The subsurface states with energy of 0.35 eV degenerated at $\overline{\Gamma}$
are localized in a few subsurface $B$ layers with 40\% of the state concentrating in 
the second $B$ layer. The third subsurface state is located at the bottom of the s bulk
boron states. 

The $Mg-t$ surface shows distinct 
electronic structure compared to the $B-t$ one. In particular,
the $Mg$ occupied surface state of the $s-p_{z}$ symmetry with energy of -1.94 eV 
appears at the $\overline{\Gamma}$ point.
Its charge density distribution localized mostly (65\%) in the $Mg$ surface layer 
and in the vacuum region is shown in Fig.4. 
The origin of this state can be understood from a simple charge transfer picture.
In bulk the $Mg$ atom donates two valence electrons to the adjacent $B$ 
planes thus moving up
to $E>E_F$ all the $Mg$ bands. In the surface layer the $Mg$ atom donates one electron to 
the subsurface $B$ plane while another electron forms the dangling bond ($s-p_z$) 
occupied surface state.

Unoccupied subsurface states with energy of 0.36 eV degenerated at $\overline{\Gamma}$ 
are formed by the subsurface $B$ 
layer, 70\% of the state is concentrated in the layer. At energy $\sim$ -12.3 eV there 
also exists a subsurface resonance state generated by the $B$ layers. 

In Fig.5 we show the calculated surface layer DOS for both the $B-t$ and $Mg-t$ surfaces
and compare them with the corresponding central layer DOS. In the $B-t$ surface 
the surface DOS at $E_F$ which also includes the vacuum region is by 55\% higher than
the central $B$ layer DOS. In the $Mg-t$ surface the surface DOS at $E_F$ is higher than 
the central $Mg$ layer DOS by factor of 2. Both these results favour the higher surface 
critical temperature $T^{s}_{c}$ compared to that in bulk. 

It is less known about
phonons on the $MgB_2(0001)$ surface. There normally exist surface phonon modes on 
metal surfaces with slightly smaller frequences compared to those in bulk \cite{hannon}.
In bulk $MgB_2$ the in-plane boron mode $E_{2g}$ is responsible for strong 
electron-phonon interaction \cite{dolgov,bohnen}. Because of its in-plane character
one can expect that the vibrational frequences and atomic displacements of this mode
in the surface or subsurface boron layer will be similar to those in bulk. Therefore
one can expect very similar or even higher $T^{s}_{c}$ compared to $T_{c}$.

Image states fall in a group of surface states which are linked to the vacuum level 
and located relatively far from the surface. The calculated work function
which fixes the vacuum level relative to $E_F$ was obtained to be of 6.1 eV for 
the $B-t$ surface and 4.2 for the $Mg-t$ one. Similar to simple and noble metal
surfaces \cite{chulkov} the wave function maximum of the n=1 image state on $MgB_2$
is located at $\sim$6 a.u. beyond the surface atomic layer for both surfaces.
In Figs.1a and 1b we show the calculated n=1,2 resonance image states. 
Nothing is known about 
the image plane position on $MgB_2$ and we varied $z_{im}$ for both terminations  
within the 2.0-3.5 a.u. interval beyond the surface layer. This variation leads
to $E_{1}=-0.9\pm 0.15$ eV and $E_{2}=-0.25\pm 0.05$ eV for the $B-t$ surface as well 
as to $E_{1}=-1.1\pm 0.15$ eV and $E_{2}=-0.30\pm 0.05 eV$ for the $Mg-t$ surface, 
the error bar includes the energy dependence on the $z_{im}$ position. The energies 
obtained are rather
similar to those for the n=1,2 resonance image states on simple metal 
surfaces \cite{chulkov}. 

The resonance image states are mostly degenerated with magnesium bulk states. 
The interaction between the image states and the Mg bulk states results in 
a different reflectivity of $B$ and $Mg$ layers and
a different behavior of the image state wave functions in bulk. The amplitude of these wave
functions is significantly larger in magnesium layers than in boron ones. This behavior
of image states is specific for $MgB_2$ due to its peculiar bulk electronic structure 
and was not found for simple and noble metals \cite{chulkov}. 

It is known that the relaxation of closed-packed simple metal surfaces
\cite{hofmann} is relatively small:
the contraction/expansion of the first interlayer spacing is $\leq$ 6\%. We have
inspected the dependence of the surface electronic structure by computing slabs 
with the contracted and expanded first interlayer spacing by 6\% for both terminations.  
We have found that these relaxations lead to change of the surface state energies  
within 0.1 eV and to small change of the surface DOS at $E_F$. The change of 
the n=1,2 image state energies is significantly smaller than the error bar. 

In conclusion, we have performed self-consistent pseudopotential calculations of 
the surface electronic structure
for the  $B-t$ and $Mg-t$ surfaces of $MgB_2$. We have found a variety of surface 
and subsurface states as well as two resonace image states on both surfaces 
including an unoccupied quantum well state 
of the $p_{x,y}$ symmetry on the $B-t$ surface.
Due to very clear surface character of these states the $MgB_2(0001)$ surfaces provide
good oportunity to test the theoretical results by measuring the surface electronic 
structure by different spectroscopies such as
photoemission, inverse photoemission, time-resolved two-photon photoemission, 
and scanning tunneling spectroscopy.
The obtained higher surface layer DOS at $E_F$  
favours the higher critical temperature compared to that in bulk. This is inconsistent with
recent STS experiments which have shown an opposite trend \cite{rubio,karapetrov,giubileo,chen,sharoni}. 
We attribute this discrepancy  
to contaminations and desorder on polycrystal sample surfaces.

We acknowledge support by the Basque Country University, Basque Hezkuntza Saila,
and Iberdrola S.A.

\begin{figure}
\caption{Projected bulk band structure of $MgB_2$ together with the surface 
and image-potential states for $B$-terminated (a) and $Mg$-terminated (b) surfaces.
The light (dark) grey areas represent the $\pi$ ($\sigma$) projected bulk
states. Dotted areas indicate the magnesium projected states. Thick solid
lines depict surface state which are mostly localized in the topmost $B$ layer.}
\end{figure}

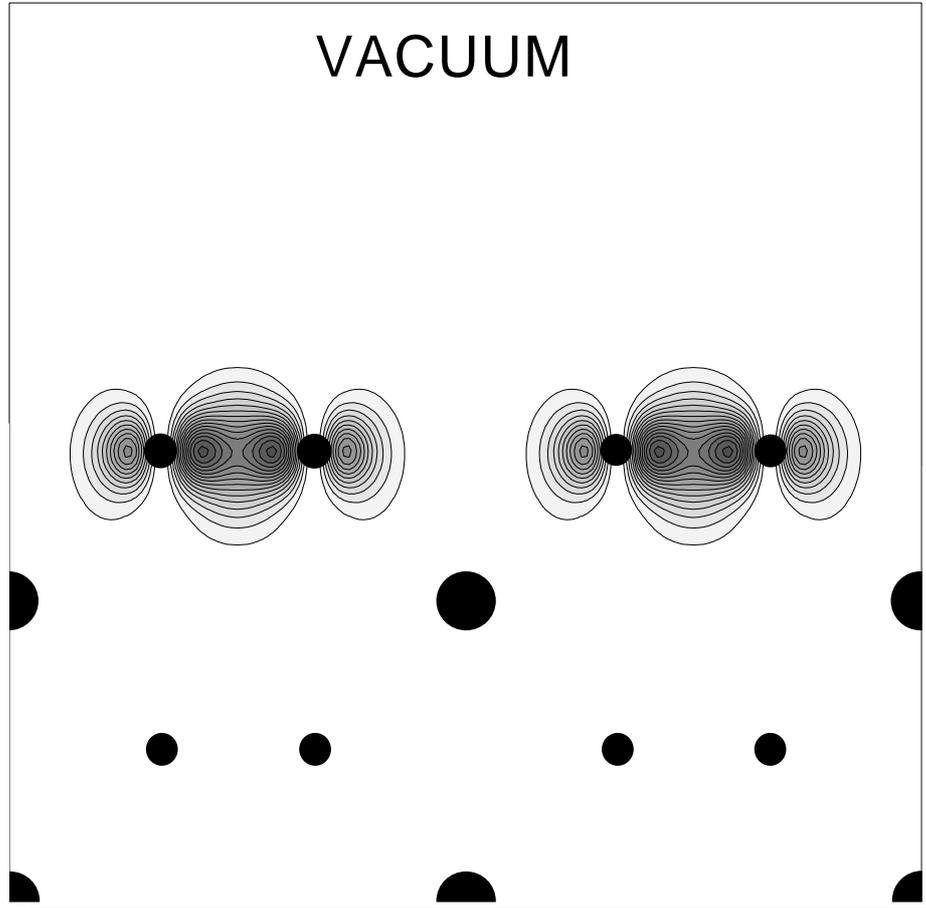
\begin{figure}
\caption{Charge density distribution of the unoccupied boron surface (quantum well)
state at $\Gamma$ in the ($10\overline{1}$0) plane for the $B$-terminated surface.
Small (large) filled circles indicate the $B$($Mg$) atom positions.}
\end{figure}

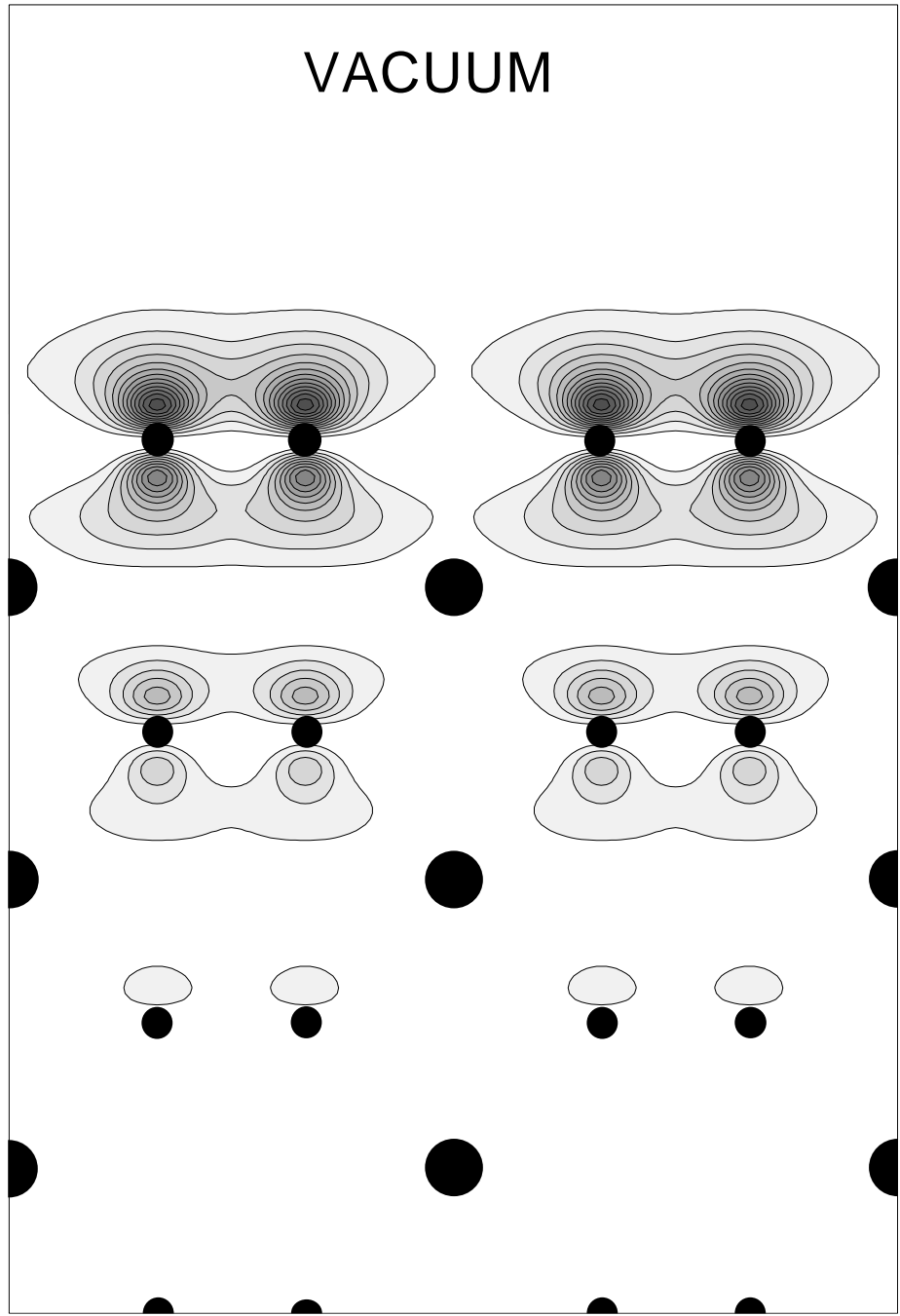
\begin{figure}
\caption{Charge density distribution of the occupied boron $p_{z}$ surface 
state at $\Gamma$ in the ($10\overline{1}$0) plane for the $B$-terminated surface.}
\end{figure}

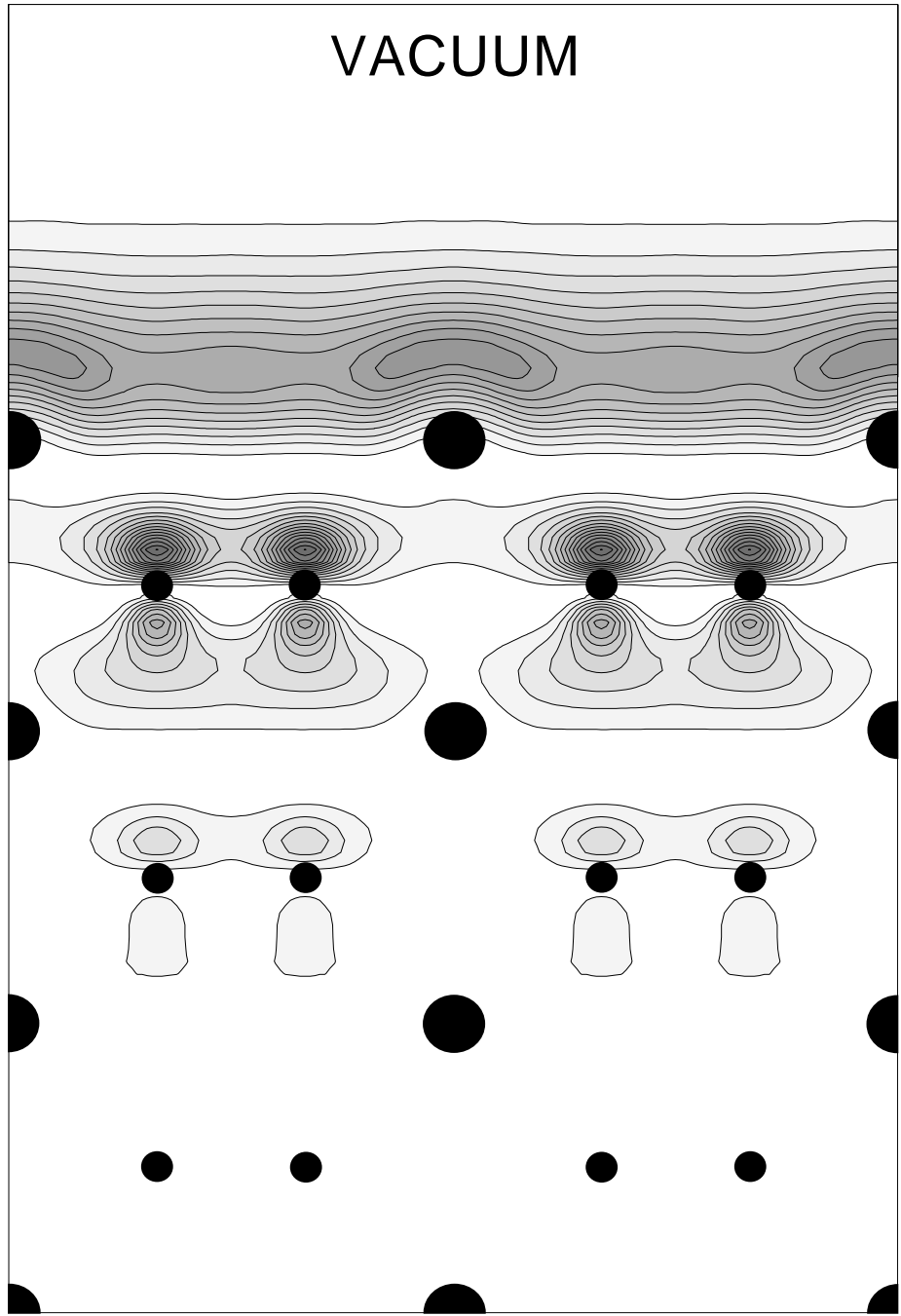
\begin{figure}
\caption{Charge density distribution of the occupied magnesium $p_{z}$ surface 
state at $\Gamma$ in the ($10\overline{1}$0) plane for the $Mg$-terminated surface.}
\end{figure}

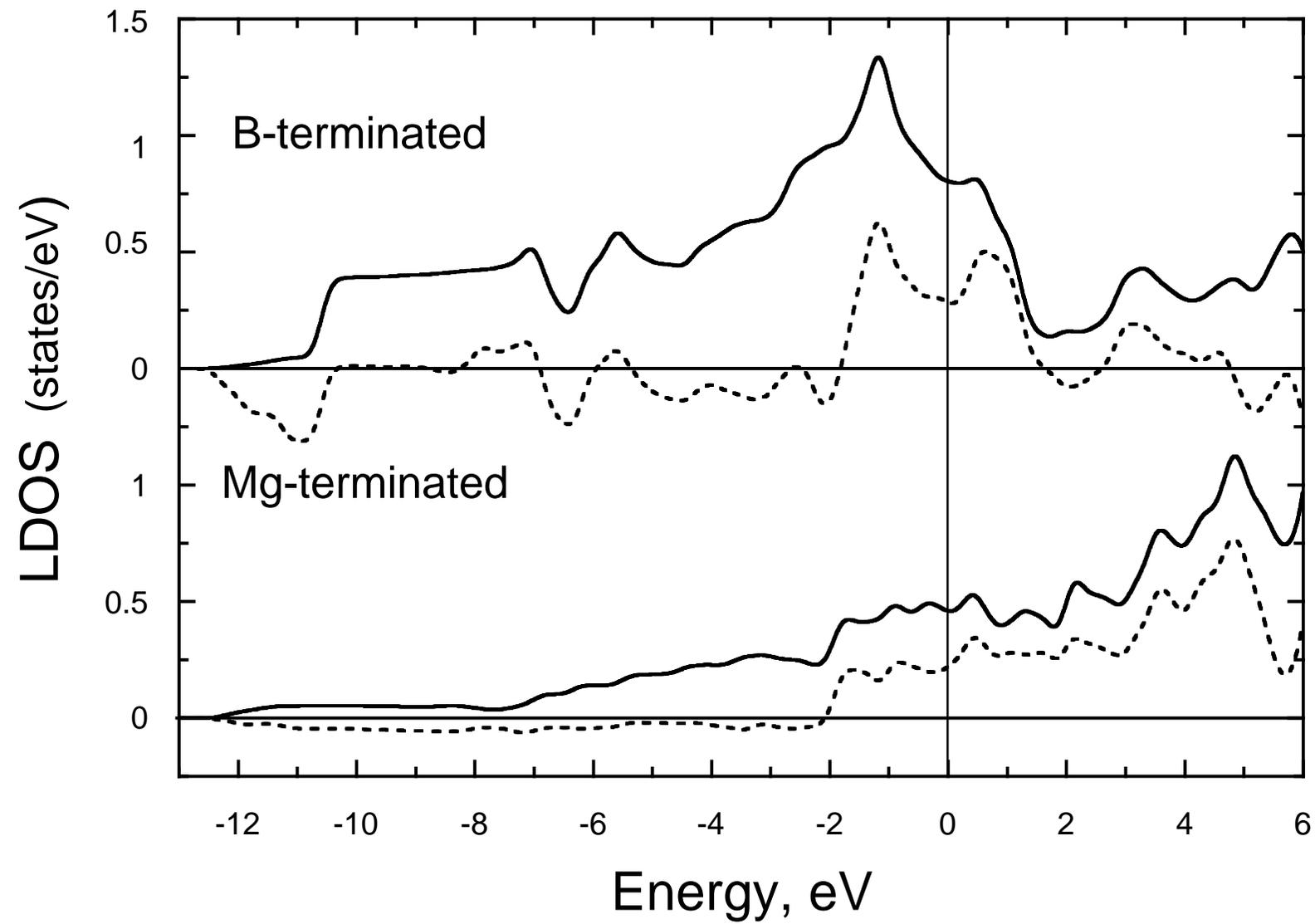
\begin{figure}
\caption{The calculated surface layer DOS at $E_F$ (solid lines) for both $B$- and
$Mg$-terminated surfaces. Dashed lines show the difference between the surface layer 
DOS and the corresponding central layer DOS.}
\end{figure}


\begin{thebibliography}{99}

\bibitem{naga2001} J. Nagamatsu {\it et al}.,
 Nature {\bf 410}, 63 (2001).

\bibitem{bud'ko} S. L. Bud'ko {\it et al}.,
 Phys. Rev. Lett. {\bf 86}, 1877 (2001).

\bibitem{finn} D. K. Finnemore  {\it et al}., 
 Phys. Rev. Lett. {\bf 86}, 2420 (2001).

\bibitem{takahashi} T. Takahashi {\it et al}., 
Phys. Rev. Lett. {\bf 86}, 4915 (2001).

\bibitem{rubio} G. Rubio-Bollinger, H. Suderow, and S.Vieira, cond-mat/0102242.

\bibitem{karapetrov} G. Karapetrov {\it et al}.,
 Phys. Rev. Lett. {\bf 86}, 4374 (2001). 

\bibitem{giubileo} F. Giubileo {\it et al}.,
cond-mat/0105146.

\bibitem{chen} C. -T. Chen {\it et al}., cond-mat/0104285.

\bibitem{sharoni} A. Sharoni, I. Felner, and O. Millo,
 Phys. Rev. B {\bf 63}, 220508R (2001). 

\bibitem{0102389} H. Schmidt {\it et al}., 
 cond-mat/0102389.

\bibitem{kortus} J. Kortus {\it et al}.,
Phys. Rev. Lett. {\bf 86}, 4656 (2001).

\bibitem{an} J. M. An and W. E. Pickett, 
Phys. Rev. Lett. {\bf 86}, 4366 (2001).

\bibitem{dolgov} Y. Kong {\it et al}.,
Phys. Rev. B {\bf 64}, 020501R (2001).

\bibitem{hirsch} J. E. Hirsch and F. Marsiglio, cond-mat/0102479

\bibitem{voelker} K. Voelker, V. I. Anisimov, and T. M. Rice, cond-mat/0103082.

\bibitem{osborn} R. Osborn {\it et al}.,
cond-mat/0103064. 

\bibitem{guinea} E. Bascones and F. Guinea, cond-mat/0103190.

\bibitem{liu} A. Y. Liu, I. I. Mazin, and J. Kortus, cond-mat/0103570.

\bibitem{gorshunov} B. Gorshunov {\it et al}., 
 cond-mat/0103164.

\bibitem{jorg} J. D. Jorgensen, D. G. Hinks, and S. Short, cond-mat/0103069.

\bibitem{saito} E. Saito {\it et al}.,
J. Phys.:Condens. Matter, {\bf 13}, L267 (2001).

\bibitem{antropov} K. D. Belashchenko, M. van Schilfgaarde, and V. P. Antropov,
con-mat/0102290.  

\bibitem{satta} G. Satta {\it et al}.,
cond-mat/0102358.

\bibitem{taku} T. J. Sato, K. Shibata, and Y. Takano, cond-mat/0102468.

\bibitem{bohnen} K. P. Bohnen, R. Heid, and B. Renker, cond-mat/0103319.

\bibitem{jung} C. U. Jung {\it et al}., cond-mat/0105330.

\bibitem{xu} M. Xu {\it et al}., cond-mat/0105271.

\bibitem{hayami} W. Hayami {\it et al}., Jpn. J. Appl. Phys. {\bf 10}, 172 (1994).

\bibitem{pesiss00} C. L. Perkins {\it et al}., 
 Surf. Sci. {\bf 470}, 215 (2000).

\bibitem{kasoprl98} H. Kawanowa {\it et al}.,
Phys. Rev. Lett. {\bf 81}, 2264 (1998).

\bibitem{description}
The Be pseudopotential is generated according to
N. Troullier and J. L. Martins, Phys. Rev. B {\bf 43}, 1993 (1991).
The exchange-correlation potential is obtained within
LDA (J. P. Perdew and A. Zunger, Phys. Rev.
B {\bf 23}, 5048 (1981)).
Convergent results for energy spectrum within 0.02 eV were obtained 
with a 25 Ry plane-wave cut-off for
the wave functions and a 18x18 special ${\bf k}_{\parallel}$ sampling points.

\bibitem{hannon} J. B. Hannon, E. J. Mele, and E. W. Plummer, Phys. Rev. B {\bf 53},
 2090 (1996).

\bibitem{chulkov} E. V. Chulkov, V. M. Silkin, and P. M. Echenique, Surf. Sci. 
{\bf 437}, 330 (1999); V. M. Silkin, E. V. Chulkov, and P. M. Echenique, Phys. 
 Rev. B {\bf 60}, 7820 (1999).

\bibitem{hofmann} Ph. Hofmann {\it et al}., Phys. Rev. B {\bf 53}, 13715 (1996).

\end{thebibliography}
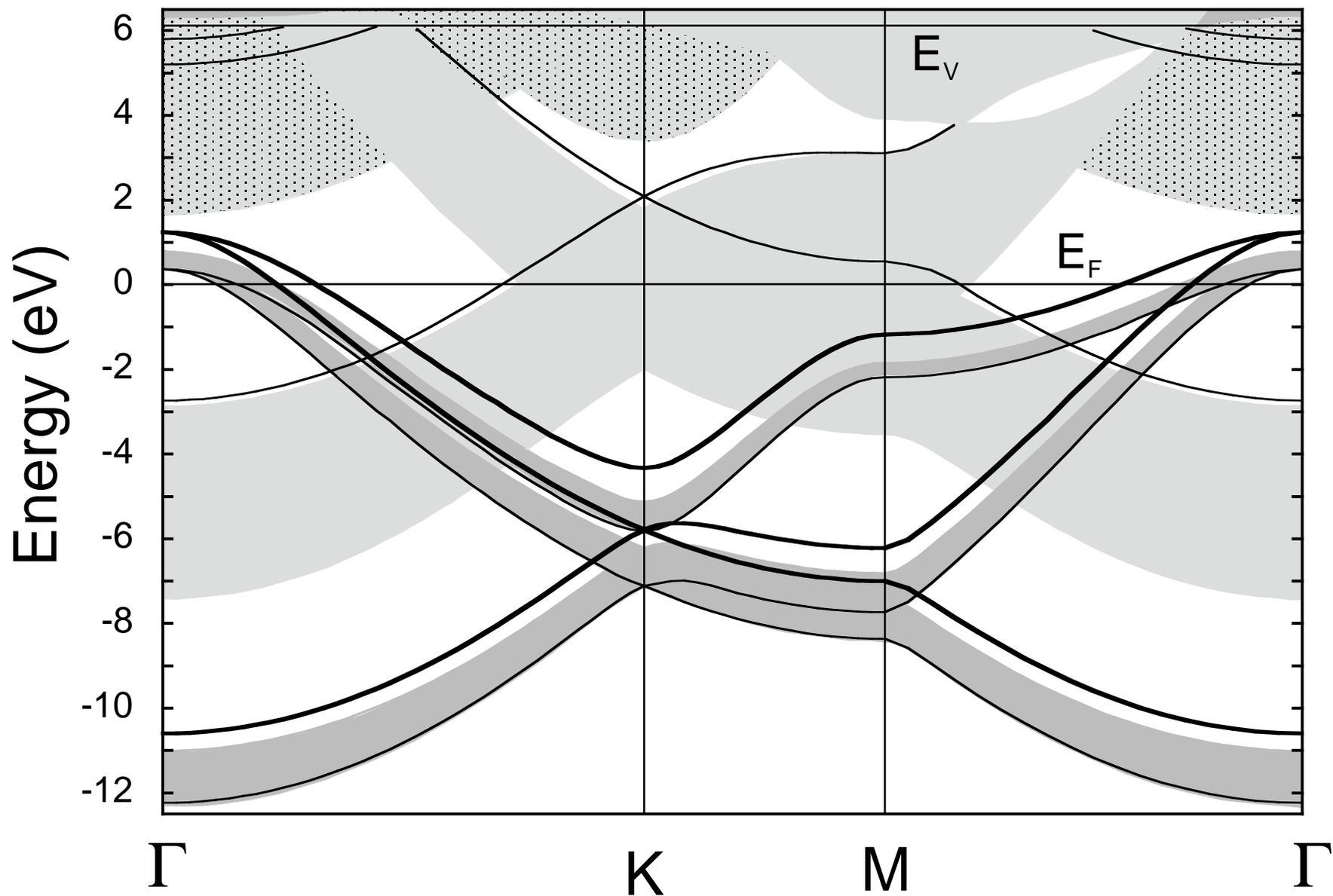
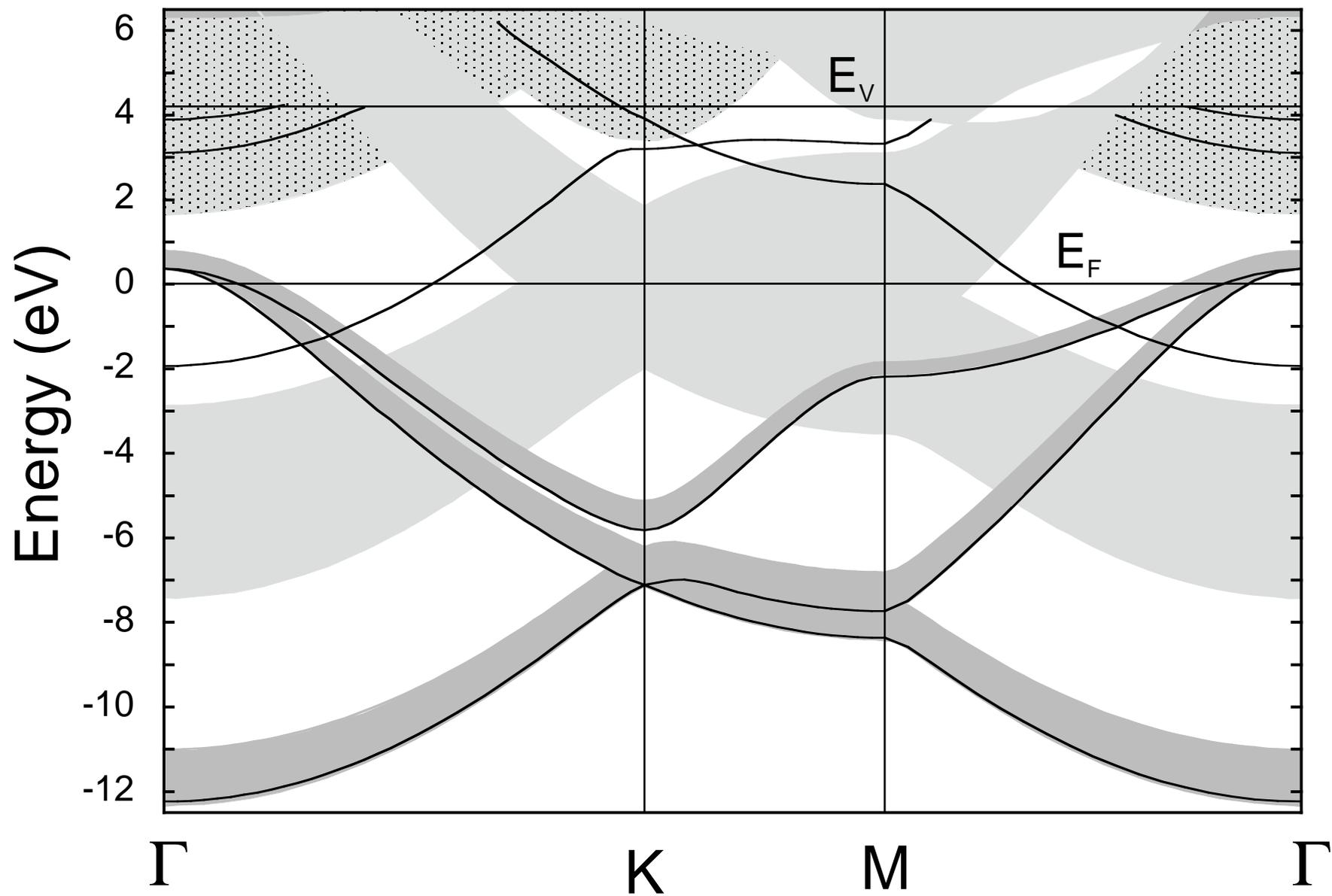
\end{document}